\documentclass[cameraready]{Interspeech}

\usepackage{cite}
\usepackage{multirow} 

\title{Bagpiper-Edit: Zero-Shot Open-Ended Audio Editing via Rich-Caption}

\author[affiliation={1,2}]{Xun}{Gong}
\author[affiliation={2}]{Jinchuan}{Tian}
\author[affiliation={1,2}]{Haoran}{Wang}
\author[affiliation={2}]{William}{Chen}
\author[affiliation={2}]{Shinji}{Watanabe}
\author[affiliation={1}]{Yanmin}{Qian}

\address{
    $^1$ Auditory Cognition and Computational Acoustics Lab, Shanghai Jiao Tong University, Shanghai, China \\
    $^2$ Language Technologies Institute, Carnegie Mellon University
}

\email{gongxun@sjtu.edu.cn}


\usepackage{comment}

\begin{document}

\maketitle

\begin{abstract}
Current text-guided audio editing methods rely on paired training data, predefined operation templates, and separate processing pipelines across speech, music, and sound.
We present Bagpiper-Edit to enable open-ended audio editing via free-form natural language instructions.
We reformulate audio editing as a rich-caption rewriting task by treating a rich caption as the semantic representation of an audio clip.
The user request is translated into an edited caption, which then guides Bagpiper-Edit to generate the target edited audio with the original audio as contextual acoustic anchor.
This unlocks the potential of free-form editing, and circumvents the need for paired audio-editing training data, enabling powerful zero-shot editing capabilities.
Evaluations across speech, audio, and free-form editing show Bagpiper-Edit maintains good consistency to the original audio and achieves similar performance to other expert models in most cases.
Demo: https://bagpiper-edit.github.io
\end{abstract}

\section{Introduction}
\setlength{\textfloatsep}{6pt}
Open-ended audio editing aims to modify an original audio recording according to a natural language user request.
Unlike text-to-audio generation, the model must perform targeted operations while keeping other content unchanged.
This requires identifying which elements to edit and which to preserve, while maintaining overall realism.
The main challenge lies in accurately interpreting the edit request in the presence of a mixture of speech, background environment, and sound events within a single waveform.

While recent Audio Large Language Models (AudioLLMs) succeed at complex audio understanding, they lack the ability to perform edits on original audio~\cite{xu2025qwen25,xu2025qwen3,kong2024audio,ghosh2025audio,ghosh2025audio3,ghosh2025musicflamingo,chu2024qwen2audio,peng2024owsm,chen2025owls,defossez2024moshi}.
Early text-guided audio editing frameworks relied on strict templates and predefined atomic operators, such as "add bird chirping" or "change to a happy voice" \cite{wang2023audit, jia2025audioeditor, xu2024prompt, manor2024zero, evans2025stable, yan2025stepaudioeditx,sioros2025editgen,copet2024simple,lam2025analyzable}.
To support open-ended user requests, recent works use LLMs to break down abstract instructions into a sequence of these basic editing steps \cite{liang2024wavcraft, peng2024voicecraft, lan2025guiding, chen2026audiochat}.
Meanwhile, other models try to achieve natural language control but are strictly limited to the speech domain, lacking the ability to modify general acoustic environments \cite{eskimez2024e2tts, yan2025minguniaudio, du2025cosyvoice3, hu2026qwen3, chen2026cosyedit, diwan2025scaling}.
While these systems attempt to support free-form inputs, they share two fundamental limitations: \textbf{(1) heavy reliance on large paired audio-editing training datasets, (2) complex compositions of multiple expert models to cover general audio editing across speech, sound effects, and music.}

\begin{figure}[t]
    \centering
    \includegraphics[width=\linewidth]{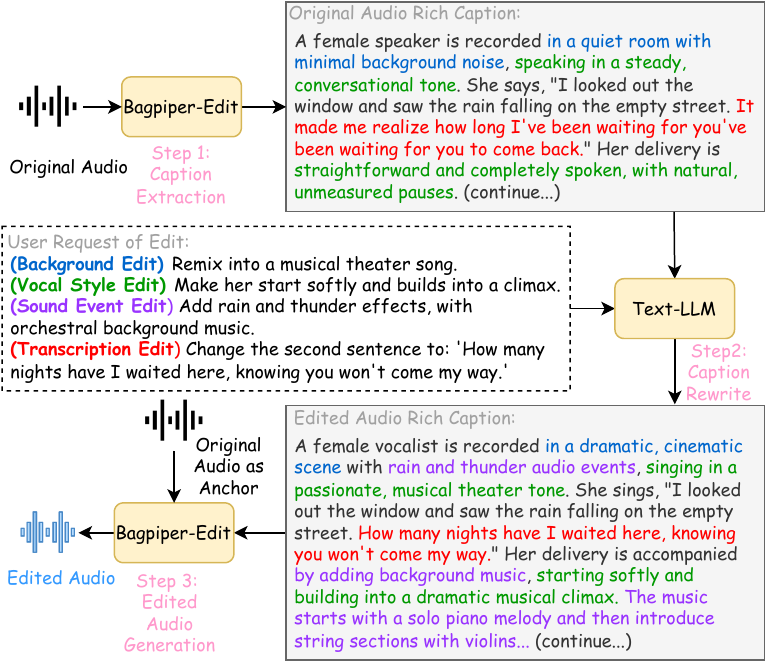}
    \caption{\textit{Bagpiper-Edit} treats a rich caption as the semantic representation of an audio clip, and defines editing as a text-space transformation before materializing the edited audio via anchored original audio.}
    \label{fig:freeform-edit}
\end{figure}

We propose \textit{Bagpiper-Edit}, a framework that reformulates open-ended audio editing as a natural language transformation.
This approach builds upon the foundational architecture of \textit{Bagpiper-Base} \cite{tian2026bagpiper}, pre-trained on caption-to-audio and audio-to-caption tasks.
Shown in Fig.\ref{fig:freeform-edit}, a \textbf{rich caption} is a detailed natural-language description of an audio clip’s events and attributes, capturing the high-level semantic (cognitive) concepts conveyed in the audio \cite{tian2026bagpiper,chen2025fusionaudio,ma2025omnicaptioner}.
However, directly using this foundation model for editing presents a challenge: the base model lacks the mechanism to handle audio-to-audio interactions.
{It tends to disregard the original audio and regenerate a new one from only the rich caption, leading to severe style and identity drift.}

To overcome these limitations, \textit{Bagpiper-Edit} upgrades the base model as a zero-shot audio editor.
As shown in Fig.\ref{fig:freeform-edit}, first, the system extracts the rich caption from the original audio.
Second, it processes a user's free-form request to rewrite this caption rather than directly modifying the signal using a strong text large language model (LLM).
Finally, it generates the new audio based on the edited caption, anchoring on the original audio.

Crucially, \textit{Bagpiper-Edit} acquires this anchoring capability \textbf{without requiring massive, expensive paired editing datasets} (i.e., original audio $\rightarrow$ edit instruction $\rightarrow$ edited audio).
Instead, we introduce a novel self-supervised training pattern, by \textit{segmenting} continuous audio into adjacent short clips and by \textit{repeating} the original audio.
Since these contiguous segments naturally share the consecutive acoustic environment, background noise, and speaker identity, we use the first segment to condition the generation of the second.
This teaches the model how to seamlessly maintain acoustic consistency across edits.

In summary, our primary contributions are:
\begin{itemize}
\item \textbf{Open-Ended Audio Editing}:
By replacing rigid edit operators with inference-time rich-caption rewriting, \textit{Bagpiper-Edit} allows users to perform complex, multi-element audio editing using intuitive, free-form natural language.

\item \textbf{Self-Supervised Zero-Shot Audio Editing}:
\textit{Bagpiper-Edit} leverages contiguous audio segments and employs an audio-repetition training strategy, thereby removing the reliance on paired edit-instruction datasets. The model naturally learns acoustic consistency, enabling powerful zero-shot editing with original audio as an effective contextual acoustic anchor.

\item \textbf{Competitive Performance Across Diverse Editing Tasks}:
Evaluations show that our zero-shot approach does not sacrifice generation quality across diverse tasks, including speech editing, sound event modification, and free-form editing. It performs on par with specialized expert models for most cases.
\end{itemize}

\section{Related work}
\label{sec:related}

\noindent \textbf{Audio Editing}:
Prior text-guided audio editing approaches often intervene directly within diffusion frameworks.
ZETA \cite{manor2024zeta} and Prompt-Guided-Edit \cite{xu2024prompt} approximate latent inversion perform text-guided editing but \cite{xu2024prompt} requires explicit token-level temporal alignments.
Alternatively, AUDIT \cite{wang2023audit} binds predefined atomic operations (such as add, remove, replace, volume up/down, etc.) directly to restricted instruction templates.
To support free-form user requests, WavCraft \cite{liang2024wavcraft},  SmartDJ \cite{lan2025guiding} and AudioChat \cite{chen2026audiochat} employ LLMs, as reasoning agents to decompose abstract requests into deterministic sequences of atomic operations.
While these pipelines improve user flexibility, they remain constrained by rigid atomic operations and the dependency on operation-specific paired training data.
In contrast, \textit{Bagpiper-Edit} bypasses these limitations by replacing atomic operations with pure text-space rich-caption rewriting.

\noindent \textbf{Speech Editing}:
Speech editing introduces strict constraints, requiring the exact preservation of transcription correctness and speaker timbre.
VoiceCraft \cite{peng2024voicecraft} treats editing as a masked reconstruction problem, heavily relying on explicit temporal alignment, while CosyEdit \cite{chen2026cosyedit} fine-tunes text-to-speech~(TTS) models on constructed edit pairs to learn the alignment.
Models like Step-Audio-EditX \cite{yan2025stepaudioeditx} rely on fixed-parameter operations to control transcription, emotion, and speaker style via synthetic paired data.
Recent systems advance the task definition by accepting natural language instructions to guide the generation process rather than requiring explicit operational commands \cite{yan2025minguniaudio, du2024cosyvoice2, du2025cosyvoice3, zhang2025mimo}.
However, besides the need for the edit-specific pair data, such specialized models struggle to generalize to complex acoustic environments: \textit{Bagpiper-Edit} overcomes this via its rich-caption abilities.

\section{Bagpiper-Edit}

\subsection{Preliminaries and Problem Formulation}
\label{sec:pre}

\textit{Bagpiper-Edit} is built upon the foundational architecture of \textit{Bagpiper-Base}~\cite{tian2026bagpiper}, a unified, autoregressive audio foundation model.
At its core, the model utilizes a decoder-only LLM to establish a bidirectional mapping between an audio signal $a$ and its rich caption $c$, thereby unifying comprehensive audio understanding ($a \rightarrow c$) and high-fidelity synthesis ($c \rightarrow a$).

For model $\theta$, an audio editing request therefore requires generating a target edited audio sequence $a'$, conditioned on the original audio $a$, and a free-form user edit request $u$,
\begin{align}
a' \sim P_\theta (\cdot | a, u). \label{eq:audio_edit}
\end{align}

\subsection{Self-Supervised Training for Acoustic Anchor}
\label{sec:ssl}

To achieve acoustic consistency without paired data, we introduce a self-supervised training paradigm.
The core idea is to train the model to maintain the acoustic environment by conditioning its generation on an acoustic anchor.

We construct training sequences consisting of two audio segments, $a_1$ and $a_2$, along with their corresponding rich captions, $c_1$ and $c_2$, using two data generation strategies:

\noindent (1) \textbf{Audio Repetition}:
The most direct way to preserve an acoustic identity is to reproduce the exact same audio.
Pairs are generated by repeating a single audio clip $a$, setting $a_1 = a_2 = a$, while the caption is $c_1 = c_2 = c$.
This fundamentally teaches the model to retain the source timbre and background environment when the caption remains unchanged.

\noindent (2) \textbf{Audio Segmentation}:
While repetition teaches exact feature matching, actual editing requires generating new content.
To achieve this, we segment a continuous audio into two adjacent clips, $a_1$ and $a_2$, and generate their respective new captions, $c_1$ and $c_2$.
Because these clips are contiguous in time, they naturally share the exact speaker identity, room acoustics, and background environment, providing ideal self-supervision for acoustic anchoring.

\begin{figure}[t]
\centering
\includegraphics[width=\linewidth]{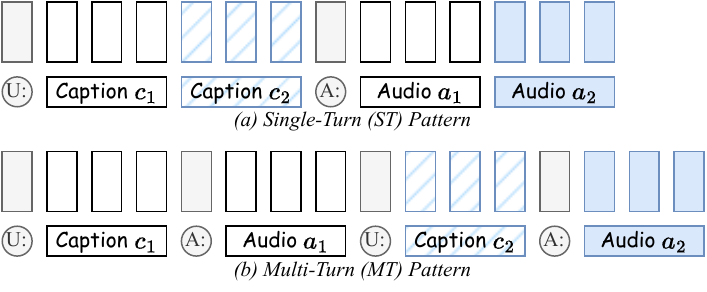}
\caption{
Dialogue patterns for \textit{Bagpiper-Edit}.
(a) The Single-Turn (\textit{ST}) pattern concatenates captions and audio within a single interaction turn. (b) The Multi-Turn (\textit{MT}) pattern structures the input as a two-round sequential dialogue.
U denotes user and A denotes assistant, both are special tags inherited from \textit{Bagpiper-Base}.}
\label{fig:patterns}
\end{figure}

We organize the constructed data into two distinct dialogue patterns for training, as illustrated in Fig.\ref{fig:patterns}.

\noindent (1) \textbf{Single-Turn (\textit{ST}) Pattern} in Fig.\ref{fig:patterns}(a) matches the format used during the pre-training phase. We concatenate the captions within a single user turn, followed by the concatenated audio in the assistant turn:
\begin{align}
    \mathrm{User}: [c_1, c_2] \rightarrow \mathrm{Assistant}: [a_1, a_2].
\end{align}
This structure provides a global semantic context before the generation of $a_2$ begins.

\noindent (2) \textbf{Multi-Turn (\textit{MT}) Pattern} in Fig.\ref{fig:patterns}(b) explicitly teaches the model audio-to-audio conditioning by formulating the task as a sequential dialogue with alternating user and assistant turns:
\begin{align}
    &\mathrm{User}: [c_{1}] \rightarrow \mathrm{Assistant}: [a_{1}], \text{ (1st Turn)} \notag \\
    &\mathrm{User}: [c_{2}] \rightarrow \mathrm{Assistant}: [a_{2}]. \text{ (2nd Turn)}
\end{align}
This sequential formulation establishes the mapping $c_1 \rightarrow a_1$ as an in-context example, helping the model maintain environmental and timbre consistency for the second turn.

\subsection{Zero-Shot Inference for Open-ended Audio Editing}
\label{sec:zero_shot}

For zero-shot audio editing from a free-form request $u$ (Eq.\ref{eq:audio_edit}), we treat $a=a_1$ as the original audio and generate the edited audio $a'=a_2$ in three steps, detailed in Fig.\ref{fig:freeform-edit}:

\noindent \textbf{Step 1: Caption Extraction}
First, we utilize \textit{Bagpiper-Edit}'s foundational audio understanding capability to extract rich caption of $a$, i.e. $c = \mathrm{Bagpiper\text{-}Edit}(a)$.

\noindent \textbf{Step 2: Caption Rewriting.} Free-form requests $u$ are typically under-specified. Instead of decomposing $u$ into atomic operations (Sec.\ref{sec:related}), we use the text LLM (Qwen3-235B-A22B-Instruct-2507-FP8)\cite{yang2025qwen3} to synthesize the target caption: $c' = \mathrm{Text LLM}(c, u)$. This seamlessly rewrites $c$ based on $u$ while retaining unchanged acoustic description text.

\noindent \textbf{Step 3: Edited Audio Generation via Acoustic Anchor.} Finally, $c'$ guides the audio generation, using $a$ as the contextual acoustic anchor:
\begin{align}
a' \sim P_{\mathrm{Bagpiper\text{-}Edit}}(\cdot \mid a, c, c'),
\end{align}
which realizes compositional edits into $a'$ in a single pass.
By resolving edits entirely in the text space, this formulation supports simultaneous complex modifications (e.g., changing transcription while adding background music), avoiding the error accumulation common in cascaded systems.


\section{Experimental setup}
\label{sec:expr_setup}

\textit{Bagpiper-Base} utilizes the Qwen3-8B-Base \cite{yang2025qwen3} and multi-stream X-Codec \cite{ye2025codec} operating at 50Hz, as the audio prediction targets, and is pre-trained on caption-to-audio synthesis, audio-to-caption understanding, and pure text modeling data~\cite{tian2026bagpiper}.

The \textit{Bagpiper-Edit} model is built based on \textit{Bagpiper-Base}, 
and we train two distinct model variants \textit{ST} and \textit{MT} based on these two patterns in Sec.\ref{sec:ssl}.
The training data consists of 500k samples, \textbf{without any paired editing data}, sampled across YODAS \cite{peng2024owsm}, LAION-Audio, Emilia-En \cite{he2024emilia}, AudioSet \cite{gemmeke2017audio}, WavCaps \cite{mei2024wavcaps}, AudioCaps \cite{kim2019audiocaps}, including speech, sound, music.
The global batch size is 128k tokens and the learning rate is $1e^{-5}$.
For inference, the same decoding strategy of \textit{Bagpiper-Base} \cite{tian2026bagpiper,ho2021classifier,hussain2025koeltts,kool2019stochastic} is followed.
The \textit{Bagpiper-Base} is decoded with \textit{MT} dialogue pattern, while \textit{Bagpiper-Edit} with two different variants.
We will release the code and evaluation scripts upon acceptance.

To comprehensively evaluate our framework, we construct three editing tasks sourced from the LibriSpeech test-clean \cite{panayotov2015librispeech} and AudioSet \cite{gemmeke2017audio} datasets: speech editing, audio-event editing, and free-form rich-caption editing.

\noindent \textbf{Speech editing}:
We evaluate speech edits to transcription, emotion, and speaking style. We measure the word error rate (WER) from whisper-large-v3 \cite{radford2023robust} and accuracy on edited spans for transcription modifications.
Speaker similarity is assessed using WavLM \cite{chen2022wavlm,anastassiou2024seedtts}, overall acoustic naturalness via DNSMOS-overall \cite{reddy2021dnsmos}, and emotional accuracy using emotion2vec \cite{ma2024emotion2vec}.

\noindent \textbf{Audio-event editing}:
The audio-event editing task focuses on the addition (i.e. original + delta = edited) and removal of target sounds (i.e. original - delta = edited = ground-truth, ground-truth is the remaining component).
To evaluate the consistency between the original audio part and the edited audio, we use VERSA~\cite{shi2025versa} to compute Fréchet Audio Distance (FAD) \cite{kilgour2019frechet} and Contrastive Language-Audio Pretraining (CLAP) scores \cite{wu2023clap}.
The CLAP scores are computed for two audio embeddings as $\mathrm{conCLAP} = \mathrm{CLAP} (a_\text{original}, a_\text{edited})$ for addition, $\mathrm{conCLAP} = \mathrm{CLAP} (a_\text{ground-truth}, a_\text{edited})$ for removal.
Furthermore, an edit CLAP score is computed to measure the success of the specific modification:
$\mathrm{editCLAP} = \mathrm{CLAP} (c_\text{delta}, a_\text{edited})$.
These metrics were found by \cite{chen2026audiochat} to highly correlate with human judgment.

\noindent \textbf{Free-form editing}:
We conduct more complex compositional changes driven entirely by natural language descriptions for sound, music, speech domains.
FAD is reported to measure the consistency between original audio and edited audio.
Caption Semantic Similarity (CapSIM) via Qwen3-Embedding-4B \cite{yang2025qwen3} is evaluated to measure semantic alignment between the reference caption (from caption rewriting) and the hypothesis caption (from caption extraction of the edited audio).

For the above tasks, we utilize Qwen3-Omni-30B-A3B-Thinking \cite{xu2025qwen3} and Gemini-3-flash\footnote{https://blog.google/products-and-platforms/products/gemini/gemini-3-flash} to score the edited audio on a scale from 1 to 5.
Please refer to the demo page for more evaluation details.

\section{Experiments}

\begin{table*}[ht]
\centering
\caption{
Speech-edit results categorized by task type: transcription change, emotion change, and speaking style change.
The overall LLM scores are judged by Qwen3-Omni-30B-A3B-Thinking.
}
\label{tab:speech_edit}
\begin{tabular}{lccc ccc cc}
\toprule
 & \multicolumn{4}{c}{Transcription Edit} & \multicolumn{3}{c}{Emotion Edit} & \multicolumn{1}{c}{Style Edit} \\
\cmidrule(lr){2-5} \cmidrule(lr){6-8} \cmidrule(lr){9-9}
Method & WER(\%)$\downarrow$ & Acc(\%)
$\uparrow$ & SpkSIM$\uparrow$ & DNSMOS$\uparrow$ & Acc\textsuperscript{2}(\%)$\uparrow$ & SpkSIM$\uparrow$ & LLM$\uparrow$ & LLM$\uparrow$ \\
\midrule
CosyVoice-3\cite{du2025cosyvoice3}
& \textbf{9.74} & \textbf{95.45} & \textbf{0.86} & \textbf{3.38}
& \textbf{12.40} & \textbf{0.74} & \textbf{3.10}
& \textbf{3.69} \\
Ming-UniAudio-Edit\cite{yan2025minguniaudio} 
& 15.79 & 66.18 & 0.84 & 3.23
& 10.89 & 0.69 & 2.50
& 3.25 \\
Step-Audio-EditX\cite{yan2025stepaudioeditx}
& 14.48 & 78.15 & 0.78 & 3.34
& 10.76 & 0.65 & 3.08
& 3.67 \\
\midrule
Bagpiper-Base\cite{tian2026bagpiper}
& 72.19 & 50.66 & 0.58 & 2.23
& 5.02 & 0.28 & 1.54
& 1.95 \\

Bagpiper-Edit (ST) 
& 19.62 & 47.11 & \textbf{0.86} & \textbf{3.26}
& 10.58 & \textbf{0.86} & 2.38
& \textbf{2.94} \\
Bagpiper-Edit (MT) 
& \textbf{14.01} & \textbf{79.76} & 0.83 & 3.15
& \textbf{11.20} & 0.84 & \textbf{2.59}
& 2.73 \\
\bottomrule
\end{tabular}
\end{table*}

\begin{table*}[t]
\centering
\caption{Audio-event editing evaluation. For the addition task, a higher editCLAP indicates edition success, for the removal task, a lower editCLAP is desired. The overall LLM scores are judged by Qwen3-Omni-30B-A3B-Thinking.}
\label{tab:audio_edit}
\begin{tabular}{l cc c c  cc c c}
\toprule
\multirow{3}{*}{Method} & \multicolumn{4}{c}{Addition} & \multicolumn{4}{c}{Removal} \\
\cmidrule(lr){2-5}\cmidrule(lr){6-9}
& \multicolumn{2}{c}{Consistency} & \multicolumn{1}{c}{Edit} & \multicolumn{1}{c}{Overall} & \multicolumn{2}{c}{Consistency} & \multicolumn{1}{c}{Edit} & \multicolumn{1}{c}{Overall} \\
\cmidrule(lr){2-3}\cmidrule(lr){4-4}\cmidrule(lr){5-5} \cmidrule(lr){6-7}\cmidrule(lr){8-8}\cmidrule(lr){9-9}
& FAD$\downarrow$ & conCLAP$\uparrow$ & editCLAP$\uparrow$ & LLM$\uparrow$
& FAD$\downarrow$ & conCLAP$\uparrow$ & editCLAP$\downarrow$ & LLM$\uparrow$ \\
\midrule
AudioLDM2\cite{liu2024audioldm2}
& 7.17 & 0.23 & 0.14 & 3.47
& 6.37 & 0.30 & -0.01 & 4.00
\\
\midrule
Bagpiper-Base\cite{tian2026bagpiper}
& 6.13 & 0.24 & 0.04 & \textbf{3.79}
& 8.36 & 0.29 & \textbf{0.07} & 4.16
\\
Bagpiper-Edit (ST) 
& \textbf{3.26} & \textbf{0.74} & 0.08 & 3.54
& 8.33 & 0.50 & 0.17 & 3.98
\\
Bagpiper-Edit (MT) 
& 3.29 & 0.51 & \textbf{0.18} & \textbf{3.79}
& \textbf{4.35} & \textbf{0.52} & \textbf{0.07} & \textbf{4.49}
\\
\bottomrule
\end{tabular}
\end{table*}

\begin{table}[t]
\centering
\caption{
Free-form rich-caption editing results.
The overall LLM scores are judged by Qwen3-Omni-30B-A3B-Thinking and Gemini-3-flash.}
\label{tab:freeform_edit}
\begin{tabular}{l cccc}
\toprule
Bagpiper & FAD$\downarrow$ & CapSIM$\uparrow$ & Qwen3$\uparrow$ & Gemini$\uparrow$ \\
\midrule
Base      & 7.62 & 0.4636 & 2.24 & 3.38 \\
Edit (ST) & \textbf{0.91} & 0.5355 & 2.60 & 3.89 \\
Edit (MT) & 2.85 & \textbf{0.5961} & \textbf{2.75} & \textbf{3.95} \\
\bottomrule
\end{tabular}
\end{table}

\subsection{Speech editing}
\label{sec:exp_speech}

\begin{figure}[t]
    \centering
    \includegraphics[width=\linewidth]{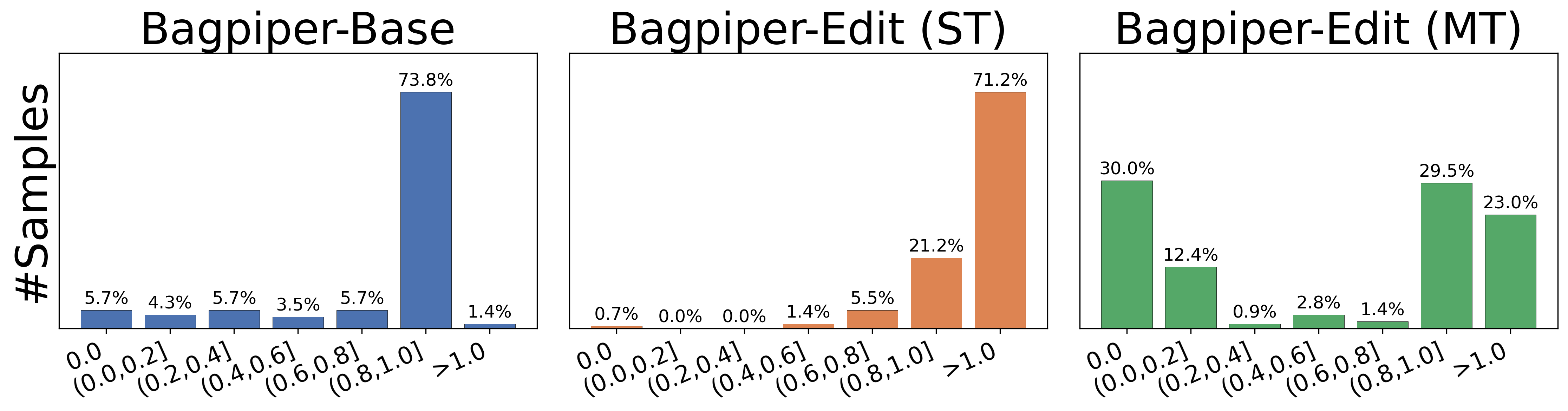}
    \caption{
    Histogram WER distributions of the full-sentence transcription replacement task (i.e. voice cloning task) for \textit{Bagpiper-Base}, \textit{Bagpiper-Edit (ST)}, \textit{Bagpiper-Edit (MT)}.
    }
    \label{fig:exp_speech_edit}
\end{figure}

We first analyze transcription editing in Tab.\ref{tab:speech_edit}.
Although provided with original audio, \textit{Bagpiper-Base} still causes severe speaker identity loss (SpkSIM = 0.58).
Our self-supervised training resolves this: the \textit{ST} pattern achieves  SpkSIM of 0.86, surpassing CosyVoice-3.
However, it also results in poor editing accuracy (47.11\%) and high failure rates in full-sentence replacements (Fig.\ref{fig:exp_speech_edit}).
Conversely, the \textit{MT} pattern yields an optimal balance, delivering a 79.76\% editing accuracy, competitive WER (14.01\%), and robust SpkSIM of 0.83.
Fig.\ref{fig:exp_speech_edit} shows that WER is largely inflated by occasional extreme failures from full-sentence replacements.
This is because the target caption for a completely new sentence is generated by the LLM rather than extracted from real audio, potentially causing errors to propagate into the synthesized audio.
While domain-specific expert models (lines 1 to 3) can mitigate these issues in strict objective metrics, they require massive paired training data.

For emotion editing task, both the \textit{ST} and \textit{MT} patterns achieve accuracy comparable to domain-specific expert models, while significantly outperforming them in SpkSIM preservation.
However, performance on speaking style editing reveals certain limitations of our base model; the scores for both patterns lag behind those of specialized models.
Notably, unlike in transcription editing, the performance gap between the \textit{ST} and \textit{MT} setups is marginal for these semantic attributes.

\subsection{Audio-event editing}
\label{sec:audio_edit}

Tab.\ref{tab:audio_edit} presents the evaluation of audio-event addition and removal tasks.
For addition, where a new sound is introduced into an existing acoustic environment, both AudioLDM2 and \textit{Bagpiper-Base} struggle to maintain background consistency, yielding poor FAD ($>6.1$) and conCLAP ($<0.25$) scores.
The \textit{ST} pattern strongly anchors to the original audio, achieving the best consistency metrics but failing to effectively insert the new event (editCLAP = 0.08).
In contrast, the \textit{MT} pattern strikes a superior balance: it successfully adds the target sound (the highest editCLAP) while maintaining the consistency, culminating in best LLM score.

The removal task, where a model must eliminate a specific sound from a mixture, is more challenging.
While AudioLDM2 achieves an optimal editCLAP, it does so at the severe expense of contextual consistency with FAD of 6.37, suggesting it synthesizes the audio using only the ground-truth caption.
The \textit{ST} pattern fails to generalize in this scenario, indicating difficulty in achieving global event rewriting, which corresponds to Fig.\ref{fig:exp_speech_edit}.
Conversely, \textit{Bagpiper-Edit (MT)} demonstrates better proficiency, successfully removes the target sound, while preserving the core acoustic background.

\footnotetext[2]{The accuracy measured by emotion2vec remains low across all systems, which is because many target emotion labels is out-of-domain with emotion2vec, and the source text's semantic content is frequently unrelated to the target emotion, significantly complicating accurate objective measurement.}

\subsection{Free-form rich-caption editing}

Tab.\ref{tab:freeform_edit} shows the evaluation of free-form rich-caption editing, challenging the models to execute complex, compositional modifications driven purely by natural language instructions.
Without self-supervised training, \textit{Bagpiper-Base} suffers from severe acoustic consistency degradation of FAD score.
\textit{Bagpiper-Edit} substantially bridges this gap.
The \textit{ST} pattern aggressively preserves the original audio, achieving the lowest FAD of 0.91, but this over-reliance limits its ability to align with complex new semantics, yielding low CapSIM.
Conversely, the \textit{MT} pattern handles complex semantic modifications more effectively, with the highest CapSIM and LLM scores while maintaining acceptable acoustic consistency.

\section{Conclusion}
This paper introduces \textit{Bagpiper-Edit}, a novel framework that reformulates open-ended audio editing as a text-space rich-caption rewriting task.
By avoiding rigid predefined operations, it provides a unified, natural language-driven interface for audio editing across speech, sound, and music domains.
A core contribution is our self-supervised training paradigm via audio segmentation and repetition.
\textit{Bagpiper-Edit} provides a strong zero-shot contextual acoustic anchor, without the need for expensive paired audio-editing datasets.
Across our benchmarks, the \textbf{Multi-Turn (\textit{MT})} setup consistently outperforms the single-turn (\textit{ST}) variant, suggesting \textit{MT} as a more reliable default under our evaluation settings.

Despite these advancements, limitations remain.
As a zero-shot model, \textit{Bagpiper-Edit} can sometimes exhibit less generation stability compared to domain-specific expert models trained on massive paired data.
Furthermore, processing highly complex acoustic environments (e.g., multi-speaker separation) remains constrained by the base model's capacity.
Future work will focus on scaling the base model's comprehension of complex scenes and exploring end-to-end alignment strategies to further streamline user interaction.

\newpage

\section{Acknowledgments}

This work was supported in part by China NSFC project under Grants No. U25A20409, and in part by SJTU Med-X (Medicine \& Engineering) Translational Research Grant (YG2025LC09).




\section{Generative AI Use Disclosure}
Generative AI tools were utilized in the preparation of this manuscript primarily for stylistic polishing and grammatical refinement; they were not used to author significant original technical content.
Additionally, as detailed in Sec.\ref{sec:expr_setup}, Large Language Models (LLMs) and multimodal foundational models were employed for evaluation.

\bibliographystyle{IEEEtran}
\bibliography{main}

\end{document}